\documentclass{JHEP3}
\usepackage{cite}
\usepackage[english]{babel}
\usepackage{amssymb,amsfonts,amsmath}
\usepackage{verbatim}
  \newcommand{\cD}{{\cal D}}

\newcommand{\cO}{{\cal O}}

  \newcommand{\cZ}{{\cal Z}}

\newcommand{\be}{\begin{equation}} \newcommand{\ee}{\end{equation}}
\newcommand{\bea}{\begin{eqnarray}} \newcommand{\eea}{\end{eqnarray}}
\newcommand{\beann}{\begin{eqnarray*}}  \newcommand{\eeann}{\end{eqnarray*}}
\newcommand{\bfig}{\begin{figure}} \newcommand{\efig}{\end{figure}}
\newcommand{\ba}{\begin{array}} \newcommand{\ea}{\end{array}}
\newcommand{\bcen}{\begin{center}} \newcommand{\ecen}{\end{center}}
\newcommand{\btab}{\begin{tabular}} \newcommand{\etab}{\end{tabular}}

\newcommand{\vev}[1]{\left\langle{#1}\right\rangle}

\newcommand{\e}{{\rm e}}

\newtheorem{Proposition}{Proposition}[section]

\newtheorem{Theorem}{Theorem}[section]
\newtheorem{Lemma}{Lemma}[section]
\newtheorem{Corrolary}{Corrolary}[section]

\newcommand{\bp}{\begin{Proposition}}   \newcommand{\ep}{\end{Proposition}}
\newcommand{\bt}{\begin{Theorem}}   \newcommand{\et}{\end{Theorem}}
\newcommand{\bl}{\begin{Lemma}}     \newcommand{\el}{\end{Lemma}}
\newcommand{\bc}{\begin{Corrolary}} \newcommand{\ec}{\end{Corrolary}}

\def\hd{\hat d}

\def\bb#1{{\pmb{#1}}}
\def\d{\partial}
\def\hd{\hat{\partial}}
\def\e{\epsilon}

\title{Ward Identities for Transport in 2+1 Dimensions}

\author{Carlos Hoyos,${}^{\dagger,1}$ Bom Soo Kim${}^{*,2}$, Yaron Oz${}^{\sharp,3}$\\
${}^\dagger$\, Department of Physics, Universidad de Oviedo,
Avda. Calvo Sotelo 18, 33007, Oviedo, Spain,\\
${}^{*}$\, Department of Physics and Astronomy,
University of Kentucky, Lexington, KY 40506, USA \\
${}^{\sharp}$\, Raymond and Beverly Sackler School of
Physics and Astronomy, Tel-Aviv University, Tel-Aviv 69978, Israel\\
E-mail: ${}^1$\,\email{hoyoscarlos@uniovi.es},${}^2$\,\email{bom.soo.kim@uky.edu},${}^3$\,\email{yaronoz@post.tau.ac.il}}

\abstract{
We use the Ward identities corresponding to general linear transformations, and  derive relations
between transport coefficients of $(2+1)$-dimensional systems.
Our analysis includes relativistic and Galilean invariant systems, as well as systems without boost invariance such as Lifshitz theories.
We consider translation invariant, as well as broken translation invariant cases, and include an external magnetic field.
Our results agree with effective theory relations of incompressible Hall fluid, and with holographic calculations in a magnetically charged
black hole background.
}

\keywords{Ward Identity, Transport Coefficients, Quantum Hall Fluid, Holography}
\preprint{FPAUO-15/02}

\begin{document}

\section{Introduction}

Field theories in 2+1 space-time dimensions describe the dynamics of a large variety of systems, ranging 
from the low energy dynamics of two-dimensional branes in string theory to many condensed matter systems such as graphene, thin films of superfluids, Quantum Hall systems and possibly high $T_c$ superconductors. Despite the fact that 
 these systems differ in their details, one can use symmetries in order to derive Ward identity relations applicable to all of them.

In a previous work \cite{Hoyos:2014lla} we used Ward identities corresponding to area preserving and conformal transformations to derive relations between the Hall conductivity and viscosity in systems with broken parity and time reversal invariance. We also derived a relation between the angular momentum density and the Hall viscosity in gapped systems. In addition, the relation is shown to be valid in the presence of massless Goldstone bosons separated by an energy gap from other excitations. 
In this work we will extend the analysis and derive relations between the shear and bulk viscosities and the conductivity. 
We will consider the Ward identities for general linear transformations, that include the area preserving transformations and dilatations as special cases. 
We will consider translation invariant, as well as broken translation invariant cases, and include an external magnetic field.
Our analysis extends the work of \cite{Bradlyn:2012ea} for Galilean invariant systems to relativistic systems, and to systems without boost invariance, such as Lifshitz theories.
In the relevant limits, our general results agree with effective theory relations of incompressible Hall fluid, and with holographic calculations in a magnetically charged
black hole background.

An outline of the paper is the following: starting from general linear transformations, we derive a Ward identity \eqref{FGeneralWI} in terms of retarded correlators and equal time commutators. In the general case, the momentum and current operators are taken to be independent, as in \S \ref{sec:gen}. After evaluating the equal time commutators, the Ward identity relates the retarded correlators to one point functions \eqref{WIGenContact}. By imposing rotation and translation invariance, we derive the Ward identity \eqref{WIGenMom}, valid for general frequency and momentum. We find agreement with previous holographic computations in magnetically charged black holes \cite{Hartnoll:2007ai,Hartnoll:2007ip,Herzog:2009xv}.  

We consider Galilean invariant systems in \S \ref{sec:Galilean}. The Ward identity that relates the retarded correlator of the current to the expectation values of charge and angular momentum is written in \eqref{WIGal1}. From the general form we derive the Ward identities \eqref{WIGalMom1}-\eqref{WIGalMom2} in the cases where the system is rotation and translation invariant.  We then consider the special cases with zero magnetic field \eqref{ZeroMagneticField} and non-zero magnetic field but small frequencies \eqref{Pisloww}. We check that our results agree with the results of \cite{Bradlyn:2012ea}. 

Systems with broken translation invariance are separately considered in \S \ref{sec:BrokenTrans}. There are more terms that contribute to the Ward identity, proportional to derivatives of one-point functions. The  Ward identity for a Galilean invariant system is given in \eqref{GIWI}. At zero magnetic field it reduces to a simple form \eqref{WIGalZeroB}  if there is rotational invariance. At zero frequency, we obtain a set of simple relations \eqref{WIGalZeroBSmallOmega} among viscosity, pressure and angular momentum. In the presence of magnetic field, the corresponding Ward identity and the relation among transport coefficients becomes more complicated as \eqref{Appen1} and \eqref{WIBBrokenTransLowFrequency}. We list the general Ward identity formulas with(out) Galilean invariance in \ref{sec:ForGWI} (\ref{sec:ForGWIG}), which are valid for the systems with rotation and time translation invariance.  

\section{The general Ward identities}

In a Galilean invariant system one can derive general relations between viscosities and conductivities by acting with time derivatives on correlators of the current \cite{Bradlyn:2012ea}. In relativistic systems, or in general systems where current and momentum are independent, the equivalent relations should be derived from correlators of the momentum (or rather the momentum density $T^{0i}$). Therefore, we will be interested in deriving relations between transport coefficients from
\begin{equation}
I^{jl}=\d_t \hd_t\left[ i\Theta(t-\hat{t})\vev{\left[ T^{0j}(t,\bb{x}),T^{0l}(\hat{t},\bb{\hat{x}})\right]}\right].
\end{equation}
Note that, before introducing any explicit form for the correlators, the relations that will be derived from the Ward identities are completely general, and will apply to any field theory, independently of whether it has an effective fluid description or not. For instance, they can also be used for states with broken translation invariance, that at low energies can behave as elastic media.

Since we will be interested also in Quantum Hall fluids, we will include in the analysis an external magnetic field. When the magnetic field is non-zero there are some subtleties that have to be taken into account. For a constant magnetic field the state is isotropic,  but the energy-momentum tensor is not conserved 
\begin{equation}\label{ConsEq}
\partial_\mu T^{\mu i}=B\e^i_{\ n} J^n.
\end{equation}
Therefore, the standard angular momentum is not conserved, $\partial_t L\neq 0$, where
\begin{equation}
L=\int d^2\bb{x}\, \e_{ij}x^i T^{0j}.
\end{equation}
Nevertheless, for a constant magnetic field it is possible to define an angular momentum operator that is conserved
\begin{equation}
 L_B= \int d^2\bb{x} \, \e_{ij}x^i\left(T^{0j}-\frac{B}{2}\e^j_{\ n}x^n J^0\right).
\end{equation}
The density $J^0$ is the generator of global $U(1)$ transformations, so the conserved angular momentum generates a combination of geometric rotations and `gauge' transformations on states. We identify
\begin{equation}\label{tB}
T_B^{0 i}=T^{0 i}-\frac{B}{2}\e^i_{ \ n} x^n J^0. 
\end{equation}
as the generators of spatial translations in the presence of a magnetic field. We have presented more details and particular examples in \cite{Hoyos:2014lla}.

We will use the following shorthand notation for energy-momentum tensor and current correlators 
\begin{eqnarray}
&G^{\mu\nu\alpha\beta}=i\Theta(t-\hat{t})\vev{\left[ T^{\mu\nu}(t,\bb{x}),T^{\alpha\beta}(\hat{t},\bb{\hat{x}})\right]},
&C^{\mu\nu\alpha\beta}=i\vev{\left[ T^{\mu\nu}(t,\bb{x}),T^{\alpha\beta}(t,\bb{\hat{x}})\right]}, \nonumber \\
&G^{\mu\nu,\alpha}=i\Theta(t-\hat{t})\vev{\left[ T^{\mu\nu}(t,\bb{x}),J^{\alpha}(\hat{t},\bb{\hat{x}})\right]},~~~
&C^{\mu\nu,\alpha}=i\vev{\left[ T^{\mu\nu}(t,\bb{x}),J^{\alpha}(t,\bb{\hat{x}})\right]},\\
&G^{\mu,\alpha\beta}=i\Theta(t-\hat{t})\vev{\left[ J^{\mu}(t,\bb{x}),T^{\alpha\beta}(\hat{t},\bb{\hat{x}})\right]},~~~
&C^{\mu,\alpha\beta}=i\vev{\left[ J^{\mu}(t,\bb{x}),T^{\alpha\beta}(t,\bb{\hat{x}})\right]}, \nonumber \\
&G^{\mu\alpha}=i\Theta(t-\hat{t})\vev{\left[ J^{\mu}(t,\bb{x}),J^{\alpha}(\hat{t},\bb{\hat{x}})\right]},~~~~~~
&C^{\mu\alpha}=i\vev{\left[ J^{\mu}(t,\bb{x}),J^{\alpha}(t,\bb{\hat{x}})\right]}. \nonumber 
\end{eqnarray}
The $G$s are retarded correlators, while the $C$s are equal time commutators. Since we are defining the correlators with a general dependence on the spatial coordinates, we will include in our analysis states with broken translation invariance. We will assume, however, that the ground state is time translation invariant. Similarly, we will use the notation $C_B$ for commutators involving the operator \eqref{tB}.
We will also use the following notation for the delta functions in time and space
\begin{equation}
 \delta=\delta(t-\hat{t}), \ \ \delta'=\delta'(t-\hat{t}), \ \ \delta_x=\delta^{(2)}(\bb{x}-\bb{\hat{x}}).
\end{equation}

One side of the Ward identity is $ I^{jl}=\d_t\hd_t G^{0j0l}$, and the other side can be obtained by taking the time derivatives and using the conservation equations \eqref{ConsEq}. We obtain 
\begin{equation}
\begin{split}
&I^{jl}
= -\delta' C^{0j0l}-\delta(\d_t-\hd_t)C^{0j0l}\\
&\quad +\d_n\hd_m G^{njml}-B\e^j_{\ n}\hd_mG^{n,ml}-B\e^l_{\ m}\d_nG^{nj,m}+B^2\e^j_{\ n}\e^l_{\ m} G^{nm}\\
&= -\delta' C^{0j0l}-\delta(\d_t-\hd_t)C^{0j0l}+\delta B\e^j_{\ n}C^{n,0l}-\delta B\e^l_{\ m}C^{0j,m}\\
&\quad +\d_n\hd_m G^{njml}+B\e^j_{\ n}\hd_tG^{n,0l}+B\e^l_{\ m}\d_tG^{0j,m}-B^2\e^j_{\ n}\e^l_{\ m} G^{nm}\\
&= -\frac{1}{2}(\d_t-\hd_t)\left(\delta C^{0j0l}\right) +\frac{1}{2}\delta\left(\d_nC^{nj0l}-\hd_mC^{0jml}\right)
+\delta \frac{B}{2} \left(\e^j_{\ n}C^{n,0l}-\e^l_{\ m}C^{0j,m}\right) \\
&\quad +\d_n\hd_m G^{njml}+B\e^j_{\ n}\hd_tG^{n,0l}+B\e^l_{\ m}\d_tG^{0j,m}-B^2\e^j_{\ n}\e^l_{\ m} G^{nm}.
\end{split}
\end{equation}
We perform a Fourier transform
\begin{equation}
\tilde{I}^{jl}=\int d(t-\hat{t}) e^{i\omega(t-\hat{t})} I^{jl}.
\end{equation}
The Ward identity can then be written as (we also use $G$ for the Fourier transform with respect to time)
\begin{equation}
\begin{split}
& \omega^2 G^{0j0l}-i\omega B\e^j_{\ n}G^{n,0l}+i\omega B\e^l_{\ m}G^{0j,m}+B^2\e^j_{\ n}\e^l_{\ m} G^{nm} \\
&=i\omega C^{0j0l} +\frac{1}{2}\left(\d_nC^{nj0l}-\hd_mC^{0jml}\right)+ \frac{B}{2}\e^j_{\ n}C^{n,0l}- \frac{B}{2}\e^l_{\ m}C^{0j,m} +\d_n\hd_m G^{njml}.
\end{split}
\end{equation}
We will make use now of the fact that $T_B$ is the density for the generator of translations, and write the following operator identity for the equal time commutators with $T_B$
\begin{equation}\label{MomOp}
i[T_B^{0i}(\bb{x}), \cO(\bb{\hat{x}})]=\d_i\cO(\bb{x})\delta_x \ ,
\end{equation}
where $\cO$ is any operator.
With this identity we can simplify the contact terms
\begin{equation}
\begin{split}
\d_n C^{nj0l} =&\d_n C_B^{nj0l}+\frac{B}{2}\e^l_{\ m}\hat{x}^m \d_nC^{nj,0}= \d_nC_B^{nj0l}=-\d_l\d_n \vev{T^{nj}}\delta_x,\\
 \hd_mC^{0jml} =& \hd_m C_B^{0jml}+\frac{B}{2}\e^j_{\ n}x^n \hd_mC^{0,ml}= \hd_mC_B^{0jml}=\d_j\d_m \vev{T^{ml}}\delta_x,
\end{split}
\end{equation}
where we used the fact that the equal time commutators with the charge density $J^0$ vanish. 

Finally, the Ward identity takes the form
\begin{equation}\label{FGeneralWI}
\begin{split}
& \omega^2 G^{0j0l}-i\omega B\e^j_{\ n}G^{n,0l}+i\omega B\e^l_{\ m}G^{0j,m}+B^2\e^j_{\ n}\e^l_{\ m} G^{nm} \\
&=i\omega C^{0j0l} + \frac{B}{2}\e^j_{\ n}C^{n,0l}- \frac{B}{2}\e^l_{\ m}C^{0j,m} +\d_n\hd_m G^{njml}\\
&\quad -\frac{1}{2}\left[\d_n\d_l \vev{T^{nj}}+\d_j \d_m \vev{T^{ml}}\right]\delta_x.
\end{split}
\end{equation}

\subsection{The general case}\label{sec:gen}

We will study first the Ward identity when the current operators $J^i$ and the momentum operators $T^{0i}$ are independent. We can simplify the contact terms (CTS) on the right hand side of equation \eqref{FGeneralWI} as follows. First, we rewrite the equal time commutator using $C_B$ with \eqref{tB}, followed by the operator identity \eqref{MomOp}, and we get 
\begin{equation}
\begin{split}
CTS &= i\omega C_B^{0j0l}+\frac{i\omega B}{2} \left[\e^j_{\ n}x^n C_B^{0,0l}+ \e^l_{\ m}\hat{x}^m C_B^{0j,0}\right] 
+ \frac{B}{2}\left[ \e^j_{\ n}C_B^{n,0l}-\e^l_{\ m}C_B^{0j,m}\right] \\
&\quad -\frac{1}{2}\left[\d_n\d_l \vev{T^{nj}}+\d_j \d_m \vev{T^{ml}}\right]\delta_x\\
&=i\omega\left[ \d_j\vev{T_B^{0l}}-\d_l \vev{T_B^{0j}}\right]\delta_x-\frac{i\omega B}{2}\left[\e^j_{\ n}x^n\d_l\vev{J^0}- \e^l_{\ m}x^m\d_j\vev{J^0}\right]\delta_x\\  
&\quad -\frac{B}{2}\left[\e^j_{\ n}\d_l \vev{J^n}+\e^l_{\ m}\d_j \vev{J^m} \right]\delta_x -\frac{1}{2}\left[\d_n\d_l \vev{T^{nj}}+\d_j \d_m \vev{T^{ml}}\right]\delta_x.
\end{split}
\end{equation}
Second, we simplify the last expression by introducing the explicit form of $T_B$ \eqref{tB} and grouping terms  
\begin{equation}\label{ctsA}
\begin{split}
CTS 
&=  i\omega\e^{jl}\left[ \e^{n}_{\ m}\d_n\vev{T^{0m}}+B\vev{J^0}\right]\delta_x \\
&\quad -\frac{1}{2}\left[\d_l (\d_n\vev{T^{nj}}+B\e^j_{\ n}\vev{J^n})+\d_j (\d_m \vev{T^{ml}}+B\e^l_{\ m}\vev{J^m})\right]\delta_x.
\end{split}
\end{equation}
If the angular momentum is non-zero and there is rotational invariance, the expectation value of the momentum density should take the following form: 
\begin{equation}\label{angmom}
\vev{T^{0i}}=\frac{1}{2}\e^{ik}\partial_k\ell_T.
\end{equation}
For a homogeneous system with a boundary, this will become a momentum density at the edge. In addition, there can be a non-zero magnetization
\begin{equation}\label{magnet}
\vev{J^{i}}=\e^{ik}\partial_k M.
\end{equation}
Introducing both expressions in \eqref{ctsA}, the Ward identity becomes
\begin{equation}\label{WIGenContact}
\begin{split}
& \omega^2 G^{0j0l}-i\omega B\e^j_{\ n}G^{n,0l}+i\omega B\e^l_{\ m}G^{0j,m}+B^2\e^j_{\ n}\e^l_{\ m} G^{nm}\\
&=  i\omega\e^{jl}\left[ -\frac{1}{2}\d^2\ell_T+B\vev{J^0}\right]\delta_x +\d_n\hd_m G^{njml}\\
&\quad -\frac{1}{2}\left[\d_l (\d_n \vev{T^{nj}}-B\d_j M)+\d_j (\d_m \vev{T^{ml}}-B\d_l M)\right]\delta_x.
\end{split}
\end{equation}
In addition to the terms that appear due to the conservation equation of the energy-momentum tensor, there are several contributions from contact terms that affect the response of the system to external sources. The terms in the last line do not contain factors of $\omega$, thus they are associated with the response to static changes in the spatial metric, {\em i.e.} to the elasticity tensor. If there is rotational invariance, the expectation value of the stress tensor will be of the form $\vev{T^{ij}}=p\delta^{ij}$. If the stress correlator is approximately a function of $\bb{x}-\bb{\hat{x}}$, such that we can perform a Fourier transform, we find the following additional contributions to the momentum-dependent correlator that are needed in order to cancel the contact terms in the Ward identity\footnote{For simplicity we consider the magnetic field to be constant.}
\begin{equation}
G^{njml}\simeq -(p-MB)\delta^{nj}\delta^{ml}.
\end{equation}
These are bulk terms: the pressure term and the inverse compressibility term $\kappa^{-1}=MB=B\frac{\partial p}{\partial B}$  that appear in the one-point function of the hydrodynamic stress tensor \cite{Jensen:2011xb,Bradlyn:2012ea,Kaminski:2014,Geracie:2014nka,Jensen:2014aa}. Note that the expectation value of the stress tensor as defined from the variation of a generating functional differs from the expectation value  of the stress tensor in flat space that we defined above, the difference being the additional term proportional to the magnetization. In order to define the viscosity tensor, these terms are subtracted from the correlator \cite{Bradlyn:2012ea,Geracie:2014nka}.

\subsubsection{Case with translation invariance}
\def\-{\!\!-\!\!}

We begin with the simplest case where the system is in a homogeneous state. In this case the expectation values of operators are
 independent of the position, and the terms with derivatives in the contact terms will vanish.
The Ward identity reduces to
\begin{equation}
\begin{split}
& \omega^2 G^{0j0l}(\omega,\bb{x}\-\bb{\hat{x}})-i\omega B\e^j_{\ n}G^{n,0l}(\omega,\bb{x}\-\bb{\hat{x}})+i\omega B\e^l_{\ m}G^{0j,m}(\omega,\bb{x}\-\bb{\hat{x}})+B^2\e^j_{\ n}\e^l_{\ m} G^{nm}(\omega,\bb{x}\-\bb{\hat{x}})\\
&=  i\omega\e^{jl}B\bar{n}\delta_x +\d_n\hd_m G^{njml}(\omega,\bb{x}\-\bb{\hat{x}}),
\end{split}
\end{equation}
where $\bar{n}=\vev{J^0}$ is the density. We Fourier transform with respect to $\bb{x}-\bb{\hat{x}}$
\begin{equation}
\begin{split}
G^{\mu\nu}(\omega,\bb{x}-\bb{\hat{x}}) =& \int \frac{d^2\bb{q}}{(2\pi)^2}e^{i\bb{q}\cdot (\bb{x}-\bb{\hat{x}})} G^{\mu\nu}(\omega,\bb{q}),\\
G^{\mu\nu\alpha\beta}(\omega,\bb{x}-\bb{\hat{x}})=&\int \frac{d^2\bb{q}}{(2\pi)^2}e^{i\bb{q}\cdot (\bb{x}-\bb{\hat{x}})} G^{\mu\nu\alpha\beta}(\omega,\bb{q}).
\end{split}
\end{equation}
Since we assume rotational invariance,
the Fourier transform of the stress tensor correlator can be expanded as
\begin{equation}\label{stressc}
\begin{split}
G^{njml}(\omega,\bb{q}) &= -i\omega\left[\eta\left(\delta^{nm}\delta^{jl}+\delta^{nl}\delta^{mj}-\delta^{nj}\delta^{ml} \right)+\zeta\delta^{nj}\delta^{ml} \right. \\
&\qquad\quad \left. +\frac{\eta_H}{2}(\e^{nm}\delta^{jl}+\e^{nl}\delta^{jm}+\e^{jm}\delta^{nl}+\e^{jl}\delta^{nm})\right],
\end{split}
\end{equation}
where $\eta$, $\zeta$ and $\eta_H$ are in general functions of $\omega$ and $\bb{q}$. 

The Fourier transform of the Ward identity reads
\begin{equation}\label{WIGenMom}
\begin{split}
& \omega^2 G^{0j0l}(\omega,\bb{q})-i\omega B\e^j_{\ n}G^{n,0l}(\omega,\bb{q})+i\omega B\e^l_{\ m}G^{0j,m}(\omega,\bb{q})+B^2\e^j_{\ n}\e^l_{\ m} G^{nm}(\omega,\bb{q})\\
&=  i\omega\e^{jl}B\bar{n} 
 -i\omega (\bb{q^2}\eta \delta^{jl} +\zeta q^j q^l+\eta_H \bb{q^2} \e^{jl}) \,.
\end{split}
\end{equation}
It is instructive to compare the Ward identity to other calculations in the literature.
We are not aware of an example where both the viscosities and the finite momentum current correlators were computed in a relativistic system.  However, we can compare with the results of Hartnoll and Kovtun \cite{Hartnoll:2007ai} at zero momentum $\bb{q^2}=0$. They computed the current correlators for the holographic dual of a magnetically charged black hole in asymptotically $AdS_4$ spacetime. In the $2+1$ dimensional field theory this implies that there is a non-zero background magnetic field turned on. According to our derivation, the leading contributions at low frequencies to the current correlators should match with the contact term $\propto B\bar{n}$.  
From their equations (48) and (50) in  \cite{Hartnoll:2007ai}, the current correlators are
\begin{equation}
G^{mn}=i\omega\frac{\bar{n}}{B}\e^{nm},\ \ G^{0n,m}=i\omega\frac{3\varepsilon}{2B} \e^{nm},\ \ G^{n,0m}=i\omega\frac{3\varepsilon}{2B} \e^{nm}.
\end{equation}
We see that $G^{nm}$ matches directly with the contact term. The expressions for $G^{0n,m}$ and $G^{n,0m}$ in the Ward identity give a term such that
\begin{equation}
 \omega^2 G^{0j0l}(\omega,\bb{0})-3\varepsilon \omega^2\delta^{jl}+B^2\e^j_{\ n}\e^l_{\ m} \left[G^{nm}(\omega,\bb{0})-i\omega\frac{\bar{n}}{B}\e^{nm}\right]+O(\omega^3)=0
\end{equation}
From the conservation equation of the energy-momentum tensor at zero momentum
\begin{equation}
G^{nm}(\omega,\bb{0})-i\omega\frac{\bar{n}}{B}\e^{nm}=\omega^2\frac{3\varepsilon}{2B^2}\delta^{nm}+O(\omega^3). 
\end{equation}
So there should be a term in $G^{0j0l}$ of the form
\begin{equation}
 G^{0j0l}(\omega,\bb{0})=\frac{3}{2}\varepsilon\delta^{jl}+O(\omega).
\end{equation}
Comparing to holographic results, equation (55) in \cite{Hartnoll:2007ip}, we see that this is what we will find by absorbing the contact terms in the energy-momentum tensor correlator. A discussion of contact terms in the holographic calculation can also be found in \cite{Herzog:2009xv}.

\subsection{Galilean invariant case}\label{sec:Galilean}

Let us compare our result with those of Bradlyn, Goldstein and Read  \cite{Bradlyn:2012ea}  in the special case where there is Galilean invariance. For a theory with Galilean invariance we impose the operator relation
\begin{equation}
T^{0i}=m J^i.
\end{equation}
Then, the Ward identity \eqref{FGeneralWI} reads
\begin{equation}
\begin{split}
&m^2\left[ \omega^2\delta^j_n\delta^l_m -i\omega \omega_c\e^j_{\ n}\delta^l_m+i\omega \omega_c\e^l_{\ m}\delta^j_n+\omega_c^2\e^j_{\ n}\e^l_{\ m} \right]G^{nm}\\
&=m^2\left( i\omega\delta^j_n\delta^l_m  + \frac{\omega_c}{2}\e^j_{\ n}\delta^l_m- \frac{\omega_c}{2}\e^l_{\ m}\delta^j_n\right)C^{nm} +\d_n\hd_m G^{njml}\\
&\quad -\frac{1}{2}\left[\d_n\d_l \vev{T^{nj}}+\d_j \d_m \vev{T^{ml}}\right]\delta_x \ ,
\end{split}
\end{equation}
where we defined the cyclotron frequency as $\omega_c=B/m$. We introduce the tensors
\begin{equation}\label{defOmega}
\Omega^i_{\ j}=i\omega\delta^i_j + \omega_c\e^i_{\ j},\ \ \bar{\Omega}^i_{\ j}=-i\omega\delta^i_j +\omega_c\e^i_{\ j}. 
\end{equation}
They have the following properties
\begin{equation}
\Omega^i_{\ k}\bar{\Omega}^k_{\ j}=(\omega^2-\omega_c^2)\delta^i_j,\ \bar{\Omega}^i_{\ k}\Omega^k_{\ j}=(\omega^2-\omega_c^2)\delta^i_j.
\end{equation}
We can rearrange the terms in the more convenient form
\begin{equation}
\begin{split}
m^2 \Omega^j_{\ n} \bar{\Omega}^l_{\ m}G^{nm}&=
\frac{m^2}{2}\left( \Omega^j_{\ n} \delta^l_m-\bar{\Omega}^l_{\ m}\delta^j_n\right)C^{nm} +\d_n\hd_m G^{njml}\\
&\quad -\frac{1}{2}\left[\d_n\d_l \vev{T^{nj}}+\d_j \d_m \vev{T^{ml}}\right]\delta_x.
\end{split}
\end{equation}

The equal time commutator gives (using $C^{00}=0$)
\begin{equation}
\begin{split}
 m^2\, C^{nm}&= C_B^{0n0m} +\frac{B}{2}\e^n_{\ s} x^s C_B^{0,0m}+\frac{B}{2}\e^m_{\ s} \hat{x}^s C_B^{0n,0}\\
&=\left[\d_n\vev{T_B^{0m}}-\d_m\vev{T_B^{0n}}-\frac{B}{2}\e^n_{\ s} x^s\d_m\vev{J^0} +\frac{B}{2}\e^m_{\ s} x^s \d_n \vev{J^0}\right]\delta_x\\
&=\left[\d_n\vev{T^{0m}}-\d_m\vev{T^{0n}}+ B\e^{nm}\vev{J^0}\right]\delta_x.
\end{split}
\end{equation}
Since we assume rotational invariance, the one-point function of the momentum density should take the same form as in \eqref{angmom}.
Thus, the equal time commutator becomes
\begin{equation}
\begin{split}
 m^2\, C^{nm}
&=\left[ \frac{1}{2}\left(\e^{ms}\d_n-\e^{ns}\d_m\right)\d_s\ell_T+\e^{nm} B\vev{J^0}\right]\delta_x\\
&=\e^{nm}\left[ -\frac{1}{2}\d^2\ell_T+ B\vev{J^0} \right]\delta_x.
\end{split}
\end{equation}
This leads to
\begin{equation}\label{WIGal1}
\begin{split}
m^2 \Omega^j_{\ n} \bar{\Omega}^l_{\ m}G^{nm}&=
(-\omega_c\delta^{jl}+i\omega\e^{jl})\left[ -\frac{1}{2}\d^2\ell_T+ B\vev{J^0} \right]\delta_x+\d_n\hd_m G^{njml}\\
&\quad -\frac{1}{2}\left[\d_n\d_l \vev{T^{nj}}+\d_j \d_m \vev{T^{ml}}\right]\delta_x.
\end{split}
\end{equation}

\subsubsection{Case with translation invariance}
The Ward identity is simplified to
\begin{equation}
\begin{split}
& m^2 \Omega^j_{\ n} \bar{\Omega}^l_{\ m}G^{nm}(\omega,\bb{x}\-\bb{\hat{x}})=
(-\omega_c\delta^{jl}+i\omega\e^{jl}) B\bar{n} \,\delta_x+\d_n\hd_m G^{njml}(\omega,\bb{x}\-\bb{\hat{x}}),
\end{split}
\end{equation}
which becomes, after a Fourier transformation in space  
\begin{equation}
\begin{split}
& m^2 \Omega^j_{\ n} \bar{\Omega}^l_{\ m}G^{nm}(\omega,\bb{q})=
(-\omega_c\delta^{jl}+i\omega\e^{jl}) B\bar{n} +q_n q_m G^{njml}(\omega,\bb{q}).
\end{split}
\end{equation}
The stress tensor correlator can be expanded as in \eqref{stressc}. This leads to
\begin{equation}
\begin{split}
m^2 \Omega^j_{\ n} \bar{\Omega}^l_{\ m}G^{nm}(\omega,\bb{q})&=
(-\omega_c\delta^{jl}+i\omega\e^{jl}) B\bar{n} -i\omega (\bb{q^2}\eta \delta^{jl} +\zeta q^j q^l+\eta_H \bb{q^2} \e^{jl})\\
&=\Omega^j_{\ n}\e^{nl} B\bar{n} -i\omega (\bb{q^2}\eta \delta^{jl} +\zeta q^j q^l+\eta_H \bb{q^2} \e^{jl}).
\end{split}
\end{equation}
If we contract with $\bar{\Omega}^i_{\ j}$ and $\Omega^k_{ \ l}$, we find
\begin{equation}\label{WIGalMom1}
\begin{split}
m^2 (\omega^2 &-\omega_c^2)^2  G^{ik}(\omega,\bb{q})=(\omega^2-\omega_c^2)(i\omega \e^{ik}+\omega_c \delta^{ik}) B\bar{n}\\
&-i\omega \bb{q}^2\bigg[\eta\left( (\omega^2+\omega_c^2)\delta^{ik}+2i\omega_c\omega \e^{ik}\right)+\eta_H  \left( (\omega^2+\omega_c^2)\e^{ik}-2i\omega_c\omega \delta^{ik}\right) \bigg. \\
&\left. \quad \qquad +\zeta\left( (\omega^2-\omega_c^2)\frac{q^i q^k}{\bb{q}^2}+\omega_c(\omega_c\delta^{ik}+i\omega \e^{ik})\right)\right].
\end{split}
\end{equation}
We identify the following form factors in the current correlator.
\begin{equation}
G^{ik}=\Pi_\delta \delta^{ik}+\Pi_{\e}\e^{ik}+\Pi_{q^2}q^i q^k+\Pi_{q^2\e}(\e^{in}q_nq^k+\e^{kn}q_n q^i).
\end{equation}
From the Ward identity they can be written in terms of the viscosities as
\begin{equation}\label{WIGalMom2}
\begin{split}
&\Pi_\delta =\frac{1}{m^2(\omega^2-\omega_c^2)}\left[\omega_c B\bar{n}-\frac{i\omega \bb{q}^2}{\omega^2-\omega_c^2}\left[\eta(\omega^2+\omega_c^2)-   2i\omega_c\omega \eta_H+\omega_c^2\zeta \right]\right],\\
&\Pi_\e =\frac{1}{m^2(\omega^2-\omega_c^2)}\left[i\omega  B\bar{n}
-\frac{i\omega \bb{q}^2}{\omega^2-\omega_c^2}\left[ 2i\omega_c\omega \eta+   (\omega^2+\omega_c^2)\eta_H+i\omega_c\omega \zeta\right]\right],\\
&\Pi_{q^2}= -\frac{i\omega\zeta}{m^2(\omega^2-\omega_c^2)},\\
&\Pi_{q^2\e}=0.
\end{split}
\end{equation}
Note that these expressions are consistent with Onsager relations $G^{ij}(B,\eta_H)=G^{ji}(-B,-\eta_H)$.

We will distinguish two different cases
\begin{itemize}
\item {\bf Zero magnetic field:}
When $B=\omega_c=0$, the form factors reduce to 
\begin{equation}\label{ZeroMagneticField}
\begin{split}
\Pi_\delta =&-\frac{i\eta \bb{q^2}}{m^2\omega},\quad
\Pi_\e =-\frac{i\eta_H \bb{q^2}}{m^2\omega},\quad 
\Pi_{q^2}= -\frac{i\zeta }{m^2\omega}.
\end{split}
\end{equation}

\item {\bf Low frequencies:}
We expand to linear order in $\omega$ (both $\zeta$ and $\eta$ may contain terms $\sim 1/\omega$)
\begin{equation}\label{Pisloww}
\begin{split}
\Pi_\delta \simeq &-\frac{\bar{n}}{m}-i\omega \frac{\bb{q^2}}{B^2}\left(\eta+\zeta \right),\\
\Pi_\e \simeq &-i\omega\frac{\bar{n}}{B}-i\omega\frac{\bb{q^2}}{B^2}\left( \eta_H+2i\frac{\omega}{\omega_c}\left( \eta+\frac{\zeta}{2}\right)\right)\\
\Pi_{q^2}\simeq & i\omega\frac{\zeta}{B^2}.
\end{split}
\end{equation}
\end{itemize}
In both cases we are in agreement with Bradlyn, Goldstein and Read \cite{Bradlyn:2012ea}. These relations hold for instance for incompressible Hall fluids, and can be derived in several effective theory descriptions \cite{Tokatly:2005,Geracie:2014nka}. 

\subsection{Cases without translation invariance}\label{sec:BrokenTrans}

When translation invariance is broken there are more terms that contribute to the Ward identity, and are present at zero magnetic field. They have at least two derivatives so na\"{\i}vely they contribute only at higher order in momentum. However, as was shown in \cite{Bradlyn:2012ea,Hoyos:2014lla} the angular momentum density $\ell_T$ in \eqref{angmom} determines the Hall viscosity in special cases (gapped and in the absence of magnetic fields).\footnote{The original relation between the two was derived in  \cite{Read:2008rn} for Galilean invariant systems using different methods, see also \cite{Tokatly2007,Tokatly2009,Read2011}.} 

The relation was derived using the algebra of linear transformations on the plane. The generators are
\begin{equation}
Q_B^{ij} =\int d^2 \bb{x}\, x^i T_B^{0j}.
\end{equation}
If the magnetic field is zero, then $T_B^{0j}$ becomes simply $T^{0j}$. The algebra is
\begin{equation}
i[Q_B^{ij},Q_B^{kl}]=\delta^{jk}Q_B^{il}-\delta^{il}Q_B^{kj}.
\end{equation}
The relation to the Ward identity we have derived previously is that the commutator of linear transformations appears in the integrated form
\begin{equation}
I^{ijkl}=\int d^2 \bb{x}\int d^2 \bb{\hat{x}} \, x^i \hat{x}^k I^{jl}(\bb{x},\bb{\hat{x}}).
\end{equation}
One can then understand the relation between the viscosity and the angular momentum because the commutator of two shear transformations is proportional to the angular momentum density:
\begin{equation}
[\sigma^1_{ij}Q_B^{ij},\sigma^3_{kl} Q_B^{kl}]\propto \e_{ij}Q_B^{ij}=L_B.
\end{equation}
Where $\sigma^1$ and $\sigma^3$ are Pauli matrices. The derivation is appealing from the point of view of physics, since it has a clear geometric interpretation that fits with the Berry phase calculations \cite{Read:2008rn,Tokatly2007,Tokatly2009,Read2011}. However, there are some subtleties in the derivation. First, the angular momentum density as defined in \eqref{angmom} is non-zero only if translation invariance is broken. Second, the integrals over space need to be regulated. Both can be related to each other, for instance if the system is in a finite volume, or confined in some way by an external potential. In order to establish a relation for the Hall viscosity of the homogeneous, infinite volume system one needs to take a limit where translation invariance is recovered. This limit does not commute with the zero momentum limit, which has to be taken first in order to fix the value of the Hall viscosity.

In this context, we want to explore if the Ward identity for linear transformations leads to more relations that will fix the zero momentum value of transport coefficients. As we have seen, in the Galilean invariant case the system is more constrained since the current and momentum correlators are not independent. We will focus on this case, but the result we will derive for zero magnetic field is also valid for relativistic systems.

For general linear transformations the Ward identity is
\begin{equation}\label{GIWI}
\begin{split}
& m^2 \int d^2\bb{x} d^2 \bb{\hat{x}} \,x^i\hat{x}^k\Omega^j_{\ n} \bar{\Omega}^l_{\ m}G^{nm}=
\Omega^j_{\ n}\e^{nl}\int d^2\bb{x} d^2 \bb{\hat{x}}  \,x^i\hat{x}^k\left[ -\frac{1}{2}\d^2\ell_T+ B\vev{J^0} \right]\delta_x\\
&\quad +\int d^2\bb{x} d^2 \bb{\hat{x}}\, x^i\hat{x}^k \d_n\hd_m G^{njml}
-\frac{1}{2}\int d^2\bb{x} d^2 \bb{\hat{x}}\, x^i\hat{x}^k\left[\d_n\d_l \vev{T^{nj}}+\d_j \d_m \vev{T^{ml}}\right]\delta_x.
\end{split}
\end{equation}
In order to match with the contact terms the stress correlator should take the form
\begin{equation}
G^{njml}(\omega,\bb{x},\bb{\hat{x}})=-i\omega \bar{\eta}^{njml}(\bb{x},\bb{\hat{x}})\delta_x+O(\omega^2).
\end{equation}
We can use that
\begin{equation}
\begin{split}
&\int d^2\bb{x} d^2 \bb{\hat{x}}\,x^i\hat{x}^k\d_n \hd_m (\bar{\eta}^{njml}\delta_x)\\
&=\int d^2\bb{x} d^2 \bb{\hat{x}}\,x^i\hat{x}^k\left[\d_n \hd_m \bar{\eta}^{njml}\delta_x+\d_n\bar{\eta}^{njml}\hd_m\delta_x+\hd_m\bar{\eta}^{njml}\d_n\delta_x+\bar{\eta}^{njml}\d_n\hd_m\delta_x\right]\\
&=\int d^2\bb{x}  \d_n \d_m (x^i x^k )\bar{\eta}^{njml}=\int d^2\bb{x}(\bar{\eta}^{ijkl}+\bar{\eta}^{kjil}),
\end{split}
\end{equation}
where we integrate by parts the derivative of the delta function and do the integrals over the delta functions. 
After doing the similar manipulations, the right hand side (RHS) of the Ward identity \eqref{GIWI} becomes
\begin{equation}
\begin{split}
RHS=&\Omega^j_{\ n}\e^{nl}\int d^2\bb{x}   \left[ -\delta^{ik}\ell_T+ x^i x^k B\vev{J^0} \right]-i\omega\int d^2\bb{x} \,(\bar{\eta}^{ijkl}+\bar{\eta}^{kjil})\\
&-\frac{1}{2}\int d^2\bb{x} \left[ \vev{T^{ij}}\delta^{kl}+\vev{T^{kj}}\delta^{il}+\vev{T^{il}}\delta^{jk}+\vev{T^{kl}}\delta^{ij}\right]+O(\omega^2).
\end{split}
\end{equation}
If there is rotational invariance $\vev{T^{ij}}=p(\bb{x^2})\delta^{ij}$, and $\vev{J^0}=n(\bb{x^2})$. Then,
\begin{equation}
\begin{split}
RHS=&
\Omega^j_{\ n}\e^{nl}\delta^{ik}\int d^2\bb{x}   \left[ -\ell_T+ \frac{\bb{x^2}}{2} B n \right]
-i\omega \int d^2\bb{x}\,(\bar{\eta}^{ijkl}+\bar{\eta}^{kjil})\\
&-(\delta^{ij}\delta^{kl}+\delta^{il}\delta^{jk})\int d^2\bb{x} \,  p+O(\omega^2).
\end{split}
\end{equation}

We expand the viscosity tensor as 
\begin{equation}
 \bar{\eta}^{ijkl}= \bar{\eta}\left(\delta^{ik}\delta^{jl}+\delta^{il}\delta^{jk}-\delta^{ij}\delta^{kl} \right)+\bar{\zeta}\delta^{ij}\delta^{kl} +\frac{\bar{\eta}_H}{2}(\e^{ik}\delta^{jl}+\e^{il}\delta^{jk}+\e^{jk}\delta^{il}+\e^{jl}\delta^{ik}).
\end{equation}
Plugging it back in the Ward identity, we get
\begin{equation}
\begin{split}
RHS=&
\Omega^j_{\ n}\e^{nl}\delta^{ik}\int d^2\bb{x}   \left[ -\ell_T+ \frac{\bb{x^2}}{2} B n \right] 
-(\delta^{ij}\delta^{kl}+\delta^{il}\delta^{jk})\int d^2\bb{x} \,  p \\
&-i\omega \int d^2\bb{x} \,\left( 2\bar{\eta}\delta^{ik}\delta^{jl}+\bar{\zeta}(\delta^{ij}\delta^{kl}+\delta^{il}\delta^{jk})+2\bar{\eta}_H\e^{jl}\delta^{ik}\right)+O(\omega^2).
\end{split}
\end{equation}
On the left hand side we have a term $\omega^2 G^{nm}$ (or $\omega^2 G^{0n0m}$ in the relativistic case in the absence of magnetic field). It is possible to argue that this is $O(\omega^2)$ or higher in frequency if there is an energy gap and no magnetic field \cite{Bradlyn:2012ea,Hoyos:2014lla}, or if the only gapless degrees of freedom are ordinary Goldstone bosons. Otherwise there can be additional contributions at low frequencies. Keeping this in mind, we will proceed to see what relations can be derived from the Ward identity.

If we set the magnetic field to zero $B=\omega_c=0$, the Ward identity becomes
\begin{equation}\label{WIGalZeroB}
\begin{split}
0=&
-i\omega\e^{jl}\delta^{ik}\int d^2\bb{x}   \ell_T 
-(\delta^{ij}\delta^{kl}+\delta^{il}\delta^{jk})\int d^2\bb{x} \,  p \\
& -i\omega \int d^2\bb{x}\,\left( 2\bar{\eta}\delta^{ik}\delta^{jl}+\bar{\zeta}(\delta^{ij}\delta^{kl}+\delta^{il}\delta^{jk})+2\bar{\eta}_H\e^{jl}\delta^{ik}\right)+O(\omega^2).
\end{split}
\end{equation}
We find the conditions at zero frequency from the Ward identities 
\begin{equation}\label{WIGalZeroBSmallOmega}
\begin{split}
 &\int d^2\bb{x} \, \bar{\eta}_H(\omega\to 0)=-\int d^2\bb{x} \,  \frac{\ell_T}{2},\\
&\int d^2\bb{x}  \, \bar{\zeta}(\omega\to 0)=\frac{i}{\omega}\int d^2\bb{x}\,   p,\\
&\lim_{\omega\to 0}\int d^2\bb{x}  \, i\omega\bar{\eta}=0.
\end{split}
\end{equation}
The first equation is the relation between the Hall viscosity and the angular momentum density. The second equation is the equivalent to the pressure term in (3.11) of \cite{Bradlyn:2012ea}, {\em i.e.} it is the {\em static} response to an external strain when the pressure is non-zero. In particular it determines the response to changes in volume. We can understand its origin using the generating functional. The logarithm of the partition function is a functional of the external sources, in particular of the metric
\begin{equation}
e^{W[g_{ij}]}=\cZ[g_{ij}]=\int \cD \Phi e^{i S[\Phi,g_{ij}]}. 
\end{equation}
Where $\Phi$ denote the fields in the theory. If we modify slightly the metric $g_{ij}\to g_{ij}+\delta g_{ij}$, the first order variation of the generating  functional is: 
\begin{equation}
\delta^{(1)} W=\int dt d^2 \bb{x}\, \sqrt{g}\vev{T^{ij}}\delta g_{ij}.
\end{equation}
The second order variation includes two kind of terms: those that come from the variation of the one-point function, and those that come from the variation of the square root of the determinant, and that correspond to changes in the volume
\begin{equation}
\delta^{(2)} W=\frac{1}{2}\int dt d^2 \bb{x}\, \sqrt{g}\left(\frac{1}{2}g^{kl}\vev{T^{ij}}+\frac{1}{2}g^{ij}\vev{T^{kl}}+\frac{\delta \vev{T^{ij}}}{\delta g_{kl}}+\frac{\delta \vev{T^{kl}}}{\delta g_{ij}}\right)\delta g_{ij}\delta g_{kl}.
\end{equation}
For $\vev{T^{ij}}=pg^{ij}$, the first kind introduce the contact term in the `bulk viscosity'.

\subsubsection{Non-zero magnetic field}  \label{sec:BrokenTransMag}

If we contract \eqref{GIWI} with $\bar{\Omega}^r_{\ j}$ and $\Omega^s_{ \ l}$, we find
\begin{equation}\label{eq1}
\begin{split}
& m^2 \int d^2\bb{x} d^2 \bb{\hat{x}} \,x^i\hat{x}^k (\omega^2-\omega_c^2)^2 G^{rs}=
(\omega^2-\omega_c^2)(i\omega \e^{rs}+\omega_c \delta^{rs})\delta^{ik}\int d^2\bb{x}   \left[ -\ell_T+\frac{\bb{x^2}}{2} B n \right]\\
&+\int d^2\bb{x} d^2 \bb{\hat{x}}\, x^i\hat{x}^k \bar{\Omega}^r_{\ j}\Omega^s_{ \ l} \d_n\hd_m G^{njml}- \bar{\Omega}^r_{\ j}\Omega^s_{ \ l} (\delta^{ij}\delta^{kl}+\delta^{il}\delta^{jk})\int d^2\bb{x} p.
\end{split}
\end{equation}
We found that the contact term proportional to $\vev{J^0}$ was also present when translation invariance is not broken. Since we are interested in the limit where translation invariance is recovered, we expect the same contributions to appear in the current correlator. Let us write the current correlator in terms of independent form factors
\begin{equation}\label{GFormFac}
G^{nm}=\left[\Pi_\delta(\bb{x},\bb{\hat{x}}) \delta^{nm}+\Pi_\e (\bb{x},\bb{\hat{x}}) \e^{nm}+\d^n\hd^m\Pi_{q^2}(\bb{x},\bb{\hat{x}})+(\e^{mo}\d^n+\e^{no}\d^m)\hd_o\Pi_{q^2\e}(\bb{x},\bb{\hat{x}})\right]\delta_x +\bar{G}^{nm}.
\end{equation}
There should be contributions that compensate the $\sim \bb{x^2}$ term in \eqref{eq1}. For configurations that vary slowly in space, we can do an expansion in derivatives such that the leading terms are
\begin{equation}
\begin{split}
\Pi_\delta =&\frac{\omega_c}{m^2(\omega^2-\omega_c^2)} B n+\d_n\hd^n \Pi_\delta^{(1)},\\
\Pi_\e =&\frac{i \omega}{m^2(\omega^2-\omega_c^2)} B n+\d_n\hd^n \Pi_\e^{(1)}.
\end{split}
\end{equation}
Then,
\begin{equation}\label{Appen1}
\begin{split}
& m^2 \int d^2\bb{x} \, (\omega^2-\omega_c^2)^2\left[2\delta^{ik}\delta^{rs}\Pi_\delta^{(1)}+2\delta^{ik}\e^{rs}\Pi_\e^{(1)}+(\delta^{ri}\delta^{ks}
+\delta^{si}\delta^{kr}) \Pi_{q^2} \right. \\
&\left. \quad +(\e^{ri}\delta^{ks}+\e^{rk}\delta^{is}+\e^{si}\delta^{rk}+\e^{sk}\delta^{ri})\Pi_{q^2\e}\right]\\
&=-(\omega^2-\omega_c^2)(i\omega \e^{rs}+\omega_c \delta^{rs})\delta^{ik}\int d^2\bb{x}  \ell_T
-\bar{\Omega}^r_{\ j}\Omega^s_{ \ l} (\delta^{ij}\delta^{kl}+\delta^{il}\delta^{jk})\int d^2\bb{x} p\\
&\quad -i\omega\int d^2\bb{x}  \bar{\Omega}^r_{\ j}\Omega^s_{ \ l} (2\bar{\eta}\delta^{ik}\delta^{jl}+\bar{\zeta}(\delta^{ij}\delta^{kl}+\delta^{il}\delta^{jk})+2\bar{\eta}_H\e^{jl}\delta^{ik}) +\cdots.
\end{split}
\end{equation}
The explicit formulas for the general Ward identities with Galilean invariance are listed in the appendix \ref{sec:ForGWI}. 

To lowest order in frequencies
\begin{equation}
\begin{split}
& m^2\int d^2\bb{x} \, \omega_c^4\left[2\delta^{ik}\delta^{rs}\Pi_\delta^{(1)}+2\delta^{ik}\e^{rs}\Pi_\e^{(1)}+(\delta^{ri}\delta^{ks}+\delta^{si}\delta^{kr}) \Pi_{q^2} \right. \\
&\quad \left. +(\e^{ri}\delta^{ks}+\e^{rk}\delta^{is}+\e^{si}\delta^{rk}+\e^{sk}\delta^{ri})\Pi_{q^2\e}\right]\\
&=
\omega_c^3 \delta^{rs}\delta^{ik}\int d^2\bb{x}  \ell_T
-\omega_c^2 (2\delta^{rs}\delta^{ik}-\delta^{rk}\delta^{si}-\delta^{sk}\delta^{ri})\int d^2\bb{x} p\\
&\quad -i\omega\omega_c^2\int d^2\bb{x} (2\bar{\eta}\delta^{ik}\delta^{rs}+\bar{\zeta}(2\delta^{rs}\delta^{ik}-\delta^{rk}\delta^{si}-\delta^{sk}\delta^{ri})
+2\bar{\eta}_H\e^{rs}\delta^{ik}).
\end{split}
\end{equation}
We get several conditions from equating the factors that multiply each of the tensor structures:
\begin{equation}\label{WIBBrokenTransLowFrequency}
\begin{split}
&m^2\omega_c^2 \int d^2\bb{x} \Pi_\delta^{(1)}(\omega\to 0)=-\int d^2\bb{x} \left(p+i\omega\bar{\zeta}+i\omega \bar{\eta}-\frac{1}{2}\omega_c\ell_T\right),\\
&m^2\omega_c^2 \int d^2\bb{x} \,\Pi_\e^{(1)}(\omega\to  0)=-\int d^2\bb{x}\, i\omega\bar{\eta}_H,\\
& m^2\omega_c^2 \int d^2\bb{x} \,  \Pi_{q^2}(\omega\to 0) = \int d^2\bb{x} (p+i\omega\bar{\zeta}),\\
&\int d^2\bb{x} \Pi_{q^2\e}(\omega\to 0)=0.
\end{split}
\end{equation}
For non-Galilean invariant cases with magnetic field, we have derived the formula for the general Ward identities in the Appendix \ref{sec:ForGWIG}.

 In the incompressible Hall fluid \cite{Bradlyn:2012ea}  the Hall viscosity is constant, and
\begin{equation}
\int d^2\bb{x}\,\bar{\zeta}=\frac{i}{\omega}\int d^2\bb{x}(p-\kappa^{-1}).
\end{equation}
Where $\kappa^{-1}=B\frac{\partial p}{\partial B}$. This will in principle give terms that contribute to the form factor $\Pi_{q^2}$ and $\Pi^{(1)}_\delta$. The values depend on the shear viscosity (or rather the shear modulus). According to  \cite{Bradlyn:2012ea,Tokatly:2005} it should vanish at zero frequency for the incompressible Hall fluid (and on general grounds for any fluid). On the other hand one expects a ``shear modulus'' from the variation of the one-point function of the stress tensor in a background metric (see e.g.~\cite{Geracie:2014nka}), if $\vev{T^{ij}}=p g^{ij}$,
\begin{equation}
\frac{\delta \vev{T^{ij}}}{\delta g_{kl}}=-pg^{ik} g^{jl}+\cdots.
\end{equation}
However, in the case where the background metric is not flat the Ward identity should be modified so derivatives become covariant derivatives. Then, in order to properly keep track of the contact terms due to the variation of the metric, the analysis should be done with a general metric. We expect that the ``shear modulus'' terms will drop from the Ward identities.

\section{Conclusions}

We have generalized the Ward identities that relate viscosities and conductivities in $2+1$ dimensions as originally derived in \cite{Bradlyn:2012ea} for systems with Galilean invariance. The Ward identity in  \eqref{WIGenContact} is completely general and valid for relativistic systems or even for systems without boost invariance, and in the presence of an external magnetic field. We have also explicitly checked that when we impose Galilean invariance our results agree with those of  \cite{Bradlyn:2012ea}.

When translation invariance is not broken the Ward identity becomes \eqref{WIGenMom}. This Ward identity can be derived from the conservation equation of the energy-momentum tensor, so it has appeared in other forms in the literature before.  We have compared with  results in a holographic model \cite{Hartnoll:2007ip,Herzog:2009xv} at zero momentum and small frequencies.  The results are consistent with each other, once contact terms in the correlators are properly taken into account. In this regard, a derivation from a generating functional may account for some of the contact terms in energy-momentum correlators more explicitly than the derivation we have presented in this paper. Notwithstanding, to the best of our knowledge the Ward identity has not been used yet to derive relations between viscosities and conductivities in a holographic model, this would be an interesting exercise for the future.

When the angular momentum or the magnetization of the system are non-zero, translation invariance is broken. The general form of the Ward identity imposes some relations between transport coefficients and these quantities. For zero magnetic fields they fix the value of the Hall viscosity and inverse compressibility at zero momentum \eqref{WIGalZeroBSmallOmega} in gapped systems. If the angular momentum is kept fixed as the breaking of translation invariance is removed, this  will fix the value of the Hall viscosity in the translationally invariant system, as was argued in \cite{Hoyos:2014lla} (see also \cite{Bradlyn:2012ea}). For gapless systems holographic calculations show that the Hall viscosity and the angular momentum are not related in general \cite{Liu:2014gto,Wu:2013vya}. It would be interesting to identify which contributions account for the difference using the general Ward identity.

\acknowledgments
We want to thank Chris Herzog for useful comments. 
This work is supported in part by the Israeli Science Foundation Center of Excellence, BSF, GIF and the I-CORE program of Planning and Budgeting Committee and the Israel Science Foundation (grant number 1937/12). This work is also partially supported by NSF Grant PHY-1214341 and the Spanish grant MINECO-13-FPA2012-35043-C02-02. C.H is supported by the Ramon y Cajal fellowship RYC-2012-10370.

\appendix 

\section{Formulas for general Ward identities} \label{sec:appendix}

\subsection{Galilean invariant case with magnetic field} \label{sec:ForGWI}

The equation \eqref{Appen1} can be written by using the expression of $\Omega^r_{\ j}$ and $\bar \Omega^s_{\ l}$  given in \eqref{defOmega} 
\begin{equation}
\begin{split}
& m^2 \int d^2\bb{x} \, (\omega^2-\omega_c^2)^2\left[\delta^{ik}\delta^{rs} (2\Pi_\delta^{(1)}+\Pi_{q^2} )  +2\delta^{ik}\e^{rs}\Pi_\e^{(1)}\right. \\
&\qquad \left. +(\delta^{ri}\delta^{ks}+\delta^{si}\delta^{kr}-\delta^{ik}\delta^{rs}) \Pi_{q^2}  +(\e^{ri}\delta^{ks}+\e^{rk}\delta^{is}+\e^{si}\delta^{rk}+\e^{sk}\delta^{ri})\Pi_{q^2\e}\right]\\
= &\int d^2\bb{x} \, \left[ - \delta^{ik} \delta^{rs} \big( \omega_c (\omega^2 - \omega_c^2) \, \ell_T   
+ (\omega^2+\omega_c^2) (p + i\omega \bar{\zeta} + 2i\omega\bar{\eta}) +4 \omega^2 \omega_c\bar{\eta}_H  \big) \right. \\
&\qquad \left.
+ \delta^{ik} \epsilon^{rs} \big(-i \omega (\omega^2 - \omega_c^2)  \, \ell_T  
 - 2i \omega \omega_c (p + i\omega \bar{\zeta}+ 2i\omega\bar{\eta}) - 2i\omega (\omega^2 + \omega_c^2) \bar{\eta}_H \big)   \right.  \\
&\qquad \left. - [(\omega^2 - \omega_c^2) (\delta^{ri} \delta^{sk} +\delta^{rk} \delta^{si}-\delta^{ik}\delta^{rs}) ](p + i\omega \bar{\zeta}) \right].
\end{split}
\end{equation}
Where we use $\delta^{ri} \epsilon^{sk} +\delta^{rk} \epsilon^{si}
-\epsilon^{ri} \delta^{sk} -\epsilon^{rk} \delta^{si}=-2 \delta^{ik} \epsilon^{rs} $.
One finds there are four independent tensor structures:  
$(\delta^{ri} \epsilon^{sk} +\delta^{rk} \epsilon^{si}
+\epsilon^{ri} \delta^{sk} +\epsilon^{rk} \delta^{si}), 
(\delta^{ri} \delta^{sk} +\delta^{rk} \delta^{si} - \delta^{ik}\delta^{rs}), \delta^{ik} \delta^{rs}$ and  
$ \delta^{ik} \epsilon^{rs} $.  Thus there are four independent Ward identities 
\begin{equation}
\begin{split}
\Pi_{q^2\e} &= 0, \\
m^2 (\omega^2- \omega_c^2)  \Pi_{q^2} &=
- (p + i\omega \bar{\zeta}), \\
m^2 (\omega^2- \omega_c^2)^2  \left[ 2\Pi_{\delta}^{(1)}+\Pi_{q^2} \right] &=
-(\omega^2 + \omega_c^2) \left(p +i\omega \bar{\zeta}+ 2i\omega\bar\eta\right) 
-\omega_c(\omega^2 - \omega_c^2)  \ell_T  -4 \omega^2\omega_c \bar \eta _H, \\
m^2 (\omega^2- \omega_c^2)^2  \Pi_{\e}^{(1)} &=
 -i \omega \omega_c  (p + i\omega \bar{\zeta}+ 2i\omega  \bar \eta ) 
-i\omega  (\omega^2 - \omega_c^2) \frac{\ell_T }{2}
-i \omega(\omega^2+ \omega_c^2) \bar \eta _H,
\end{split}
\end{equation}
where the integral $ \int d^2 {\bb{x}} $ is assumed for both sides of the equations. Note that Onsager relations are satisfied.
These are the general ward identities that should be satisfied for a Galilean invariant system with time translation and 
rotation invariance (not necessarily space translation invariance) in the presence of a constant magnetic field. 

\subsection{General case with magnetic field} \label{sec:ForGWIG}

Starting from the general Ward identity \eqref{FGeneralWI}, we obtain 
\begin{equation}\label{XspaceGenWI}
\begin{split}
& \int d^2\bb{x} d^2 \bb{\hat{x}} \,x^i\hat{x}^k \left(\omega^2 G^{0j0l}-i\omega B\e^j_{\ n}G^{n,0l}+i\omega B\e^l_{\ m}G^{0j,m}+B^2\e^j_{\ n}\e^l_{\ m} G^{nm}\right) \\
&=-i\omega \e^{jl} \delta^{ik}\int d^2\bb{x}   \left[ \ell_T-\frac{\bb{x^2}}{2} B n \right]
+\int d^2\bb{x} d^2 \bb{\hat{x}}\, x^i\hat{x}^k \d_n\hd_m G^{njml}- (\delta^{ij}\delta^{kl}+\delta^{il}\delta^{jk})\int d^2\bb{x} p.
\end{split}
\end{equation}
We would like to evaluate this Ward identity by using similar form given in \eqref{GFormFac}
\begin{equation}
\begin{split}
G^{0n,m}&=\left[\Pi^{TJ}_\delta \delta^{nm}+\Pi^{TJ}_\e  \e^{nm}+\d^n\hd^m\Pi^{TJ}_{q^2}+(\e^{mo}\d^n+\e^{no}\d^m)\hd_o\Pi^{TJ}_{q^2\e}\right]\delta_x +\bar{G}^{0j,m}, \\
G^{n,0m}&=\left[\Pi^{JT}_\delta \delta^{nm}+\Pi^{JT}_\e\e^{nm}+\d^n\hd^m\Pi^{JT}_{q^2}
+(\e^{mo}\d^n+\e^{no}\d^m)\hd_o\Pi^{JT}_{q^2\e}\right]\delta_x +\bar{G}^{0n,m}, \\
G^{0n,0m}&=\left[\Pi^{TT}_\delta \delta^{nm}+\Pi^{TT}_\e  \e^{nm}+\d^n\hd^m\Pi^{TT}_{q^2}
+(\e^{mo}\d^n+\e^{no}\d^m)\hd_o\Pi^{TT}_{q^2\e}\right]\delta_x +\bar{G}^{0n,0m},  
\end{split}
\end{equation}
along with
\begin{equation}
\begin{split}
\Pi_\delta =&\Pi_\delta^{(0)}+\d_n\hd^n \Pi_\delta^{(1)},\\
\Pi_\e =&\Pi_\e^{(0)}+\d_n\hd^n \Pi_\e^{(1)}.
\end{split}
\end{equation}
Where we use shorthand notation $ \Pi = \Pi (\bb{x},\bb{\hat{x}})$. 

The general Ward identity \eqref{XspaceGenWI} becomes 
\begin{equation}
\begin{split}
& \int d^2\bb{x} \, \frac{\bb{x}^2}{2} \left[\delta^{ik}\delta^{jl} (\omega^2 \Pi^{TT(0)}_\delta 
+i\omega B [\Pi^{JT(0)}_\e + \Pi^{TJ(0)}_\e] + B^2 \Pi^{JJ(0)}_\delta) \right. \\
&\qquad  \left. +\delta^{ik}\e^{jl} (\omega^2 \Pi^{TT(0)}_\delta 
+i\omega B [\Pi^{JT(0)}_\e + \Pi^{TJ(0)}_\e] + B^2 \Pi^{JJ(0)}_\delta) \right] \\
&+\int d^2\bb{x} \,  \left[(\e^{jk}\delta^{li}+\e^{ji}\delta^{kl}+\e^{lk}\delta^{ij}+\e^{li}\delta^{jk}) 
(\omega^2 \Pi^{TT}_{q^2\e} - B^2 \Pi^{JJ}_{q^2\e} - \frac{i\omega B}{2} [\Pi^{JT}_{q^2}-\Pi^{TJ}_{q^2}] ) \right. \\
&\qquad +(\delta^{ij}\delta^{kl}+\delta^{il}\delta^{jk}-\delta^{ki}\delta^{jl}) 
(\omega^2 \Pi^{TT}_{q^2} +2i\omega B[\Pi^{JT}_{q^2\e} - \Pi^{TJ}_{q^2\e}] -B^2 \Pi^{JJ}_{q^2})  \\
&\qquad +\delta^{ki}\delta^{jl} 
(\omega^2 [\Pi^{TT}_{q^2}+2 \Pi^{TT(1)}_{\delta}]+B^2 [\Pi^{JJ}_{q^2} +2 \Pi_\delta^{JJ(1)}] 
+2i\omega B [\Pi^{JT(1)}_\e + \Pi^{TJ(1)}_{\e}]  ) \\
&\qquad  \left. +\delta^{ik}\e^{jl} (2\omega^2 \Pi^{TT(1)}_\e 
-2i\omega B [\Pi^{JT(1)}_\delta + \Pi^{TJ(1)}_\delta ]
+ 2B^2 \Pi^{JJ(1)}_\e - i\omega B [\Pi^{JT}_{q^2}+\Pi^{TJ}_{q^2}]) \right] \\
&= \int d^2\bb{x} \, \left[ -i\omega  \delta^{ik}\e^{jl} \left(\ell_T -\frac{\bb{x}^2}{2} Bn \right) 
-(\delta^{ij}\delta^{kl}+\delta^{il}\delta^{jk}-\delta^{ki}\delta^{jl})(p +  i\omega \bar \zeta )
\right. \\
&\qquad  \Big. + \delta^{ki}\delta^{jl} (p + i\omega \bar \zeta + 2i \omega \bar \eta )
- 2i \omega \delta^{ik}\e^{jl}  \bar{\eta}_H \bigg].
\end{split}
\end{equation}
By comparing the tensor structure of both sides as before, 
$(\e^{jk}\delta^{li}+\e^{ji}\delta^{kl}+\e^{lk}\delta^{ij}+\e^{li}\delta^{jk}) , 
(\delta^{ij}\delta^{kl}+\delta^{il}\delta^{jk}-\delta^{ki}\delta^{jl}), 
\delta^{ik}\e^{jl}, \delta^{ik}\e^{jl} $, we get 
\begin{equation}\label{GWi1}
\begin{split}
&\int d^2\bb{x} \, \frac{\bb{x}^2}{2} \left[ \omega^2 \Pi^{TT(0)}_\delta 
+i\omega B [\Pi^{JT(0)}_\e + \Pi^{TJ(0)}_\e] + B^2 \Pi^{JJ(0)}_\delta \right] =0 , \\
&\int d^2\bb{x} \, \frac{\bb{x}^2}{2} \left[ \omega^2 \Pi^{TT(0)}_\delta 
+i\omega B [\Pi^{JT(0)}_\e + \Pi^{TJ(0)}_\e] + B^2 \Pi^{JJ(0)}_\delta \right] =i\omega 
\int d^2\bb{x} \, \frac{\bb{x}^2}{2}  Bn , 
\end{split}
\end{equation} 
for the terms with additional factor $\bb{x}^2$ inside the integral, and the remaining parts 
\begin{equation}\label{GWi2}
\begin{split}
&\omega^2 \Pi^{TT}_{q^2\e} - B^2 \Pi^{JJ}_{q^2\e} - \frac{i\omega B}{2} [\Pi^{JT}_{q^2}-\Pi^{TJ}_{q^2}] =0 , \\
&\omega^2 \Pi^{TT}_{q^2} +2i\omega B[\Pi^{JT}_{q^2\e} - \Pi^{TJ}_{q^2\e}] -B^2 \Pi^{JJ}_{q^2}=-(p+ i\omega \bar \zeta) , \\
&\omega^2 [\Pi^{TT}_{q^2}+2 \Pi^{TT(1)}_{\delta}]+B^2 [\Pi^{JJ}_{q^2} +2 \Pi_\delta^{JJ(1)}] 
+2i\omega B [\Pi^{JT(1)}_\e + \Pi^{TJ(1)}_{\e}] =p +i\omega \bar \zeta + 2i\omega \bar \eta , \\
&2\omega^2 \Pi^{TT(1)}_\e -2i\omega B [\Pi^{JT(1)}_\delta + \Pi^{TJ(1)}_\delta ]
+ 2B^2 \Pi^{JJ(1)}_\e - i\omega B [\Pi^{JT}_{q^2}+\Pi^{TJ}_{q^2}]= -i\omega \ell_T - 2i \omega \bar \eta_H .  
\end{split}
\end{equation} 
where we suppress the integral $ \int d^2 {\bb{x}} $. 
The general ward identities, equations \eqref{GWi1} and \eqref{GWi2}, are valid without Galilean invariance. 
Only the time translation and rotational invariances are required.

\end{document}